\documentclass[sigconf]{acmart}
\AtBeginDocument{%
  }
 \renewcommand\footnotetextcopyrightpermission[1]{} % removes footnote with conference information in first column
\usepackage{booktabs}
\usepackage{multirow}
\usepackage{graphicx}
\usepackage[normalem]{ulem}
\usepackage{tabularx} % 用于自动调整列宽
\usepackage[most]{tcolorbox}
\newcommand{\yun}[1]{\textcolor{black}{#1}}

\newcommand{\wrq}[1]{\textcolor{black}{#1}}

\begin{document}

%%
%% The "title" command has an optional parameter,
%% allowing the author to define a "short title" to be used in page headers.
\title{On the Effectiveness of Context Compression for Repository-Level Tasks: An Empirical Investigation}

%%
%% The "author" command and its associated commands are used to define
%% the authors and their affiliations.
%% Of note is the shared affiliation of the first two authors, and the
%% "authornote" and "authornotemark" commands
%% used to denote shared contribution to the research.
% \author{Ben Trovato}
% \authornote{Both authors contributed equally to this research.}
% \email{trovato@corporation.com}
% \orcid{1234-5678-9012}
% \author{G.K.M. Tobin}
% \authornotemark[1]
% \email{webmaster@marysville-ohio.com}
% \affiliation{%
%   \institution{Institute for Clarity in Documentation}
%   \city{Dublin}
%   \state{Ohio}
%   \country{USA}
% }

\author{Jia Feng}
\affiliation{%
  \institution{Harbin Institute of Technology}
  \city{Shenzhen}
  \country{China}
  }
\email{jiafeng@stu.hit.edu.cn}

\author{Zhanyue Qin}
\affiliation{%
  \institution{Harbin Institute of Technology}
  \city{Shenzhen}
  \country{China}
  }
\email{johnneyqin@gmail.com}

\author{Cuiyun Gao}
\authornote{Corresponding author}
\affiliation{%
  \institution{Harbin Institute of Technology}
  \city{Shenzhen}
  \country{China}
  }
\email{gaocuiyun@hit.edu.cn}

\author{Ruiqi Wang}
\affiliation{%
  \institution{Harbin Institute of Technology}
  \city{Shenzhen}
  \country{China}
  }
\email{24s151158@stu.hit.edu.cn}

\author{Chaozheng Wang}
\affiliation{%
  \institution{The Chinese University of Hong Kong}
  \city{Hong Kong}
  \country{China}
  }
\email{adf111178@gmail.com}

\author{Yingwei Ma}
\affiliation{%
  \institution{Independent Researcher}
  \city{Beijing}
  \country{China}
  }
\email{yingwei.ywma@gmail.com}

\author{Xiaoyuan Xie}
\affiliation{%
  \institution{Wuhan University}
  \city{Wuhan}
  \country{China}
  }
\email{xxie@whu.edu.cn}

%%
%% By default, the full list of authors will be used in the page
%% headers. Often, this list is too long, and will overlap
%% other information printed in the page headers. This command allows
%% the author to define a more concise list
%% of authors' names for this purpose.
\renewcommand{\shortauthors}{Trovato et al.}

%%
%% The abstract is a short summary of the work to be presented in the
%% article.
\begin{abstract}
Repository-level code intelligence tasks, such as cross-file completion and project-aware code generation, require large language models (LLMs) to process long, multi-file contexts spanning complex dependencies. \wrq{Such lengthy} inputs introduce three practical challenges: \wrq{crucial context} can be obscured by surrounding noise across multiple files, \wrq{or be truncated due to the} limited context windows, and inference latency and memory consumption grow rapidly with context length. Context compression mitigates these \wrq{risks} by condensing long inputs into compact surrogates. \wrq{While it} has received considerable attention in the natural language processing literature\wrq{,} its applicability to code tasks remains largely unexplored; beyond efficiency gains, compression may further benefit LLM \wrq{performance} on code tasks by reducing contextual noise inherent in large, multi-file repositories.

% We present the first systematic empirical study of context compression for repository-level code intelligence. We organize eight representative \yun{context compression} methods by their output representation space into three categories, \yun{including} compressing context into discrete texts, continuous latent vectors, or visual tokens, and evaluate them on code generation and completion, two of the most widely studied tasks in repository-level code intelligence, across two(\qin{qin: multiple}) model variants (\qin{qin: the 3B and 7B versions of Qwen2.5-Coder and Qwen2.5-VL}) \yun{(3B to 7B)}. Our results show that compression need not sacrifice performance: latent-vector compression surpasses full-context performance by up to 28\% on completion tasks, suggesting its learned bottleneck acts as a task-conditioned denoiser that amplifies task-relevant signals rather than simply discarding tokens. On the efficiency side, visual-based compression reduces inference latency by up to 38\% at high compression ratios while maintaining negligible preprocessing overhead, converging to the resource cost of context-free inference. 
% These findings provide actionable guidance for selecting compression strategies under diverse task requirements and deployment constraints.
\wrq{Therefore, we} present the first systematic empirical study of context compression for repository-level code intelligence, organizing eight representative methods into three paradigms by their output representation space: discrete token sequences, continuous latent vectors, and visual tokens. We evaluate \wrq{them} on code completion and generation, two of the most widely studied tasks in repository-level code intelligence, jointly measuring task performance and deployment efficiency. Our results demonstrate that context compression is effective for code: at 4$\times$ compression, methods based on continuous latent vectors surpass full-context performance by up to 28.3\% \yun{with respect to the} BLEU \yun{score}, indicating that the latent \wrq{vector compression} filters repository noise rather than merely truncating context. 
% On the efficiency side, visual compression achieves up to a 33\% latency reduction over full context while converging to \wrq{\sout{near-plain-level} the} resource cost \wrq{of context-free prompts} at high compression ratios. 
 % On the efficiency side, all three paradigms substantially reduce inference cost relative to full-context decoding: visual compression reduces end-to-end latency by up to 50\% at high compression ratios, converging to the resource cost of inference without repository context; latent-vector compression delivers consistent reductions of up to 37\% in latency and 40\% in GPU memory across all evaluated ratios; and text-based compression achieves up to 50\% latency reduction at its most aggressive settings.
 On the efficiency side, all three paradigms reduce inference cost relative to full-context decoding. In particular, both visual and text-based compression achieve up to 50\% reduction in end-to-end latency at high compression ratios, approaching the cost of inference without repository context.
These findings establish context compression as a viable approach for repository-level code intelligence and provide guidance for paradigm selection under different deployment constraints.
\end{abstract}

%%
%% The code below is generated by the tool at http://dl.acm.org/ccs.cfm.
%% Please copy and paste the code instead of the example below.
%%

%%
%% Keywords. The author(s) should pick words that accurately describe
%% the work being presented. Separate the keywords with commas.
% \keywords{Do, Not, Use, This, Code, Put, the, Correct, Terms, for,
%   Your, Paper}
%% A "teaser" image appears between the author and affiliation
%% information and the body of the document, and typically spans the
%% page.

%%
%% This command processes the author and affiliation and title
%% information and builds the first part of the formatted document.
\maketitle

\section{Introduction}

Large language models (LLMs) have \wrq{revolutionized many} software engineering \wrq{tasks,} including code generation, program repair, code summarization, and automated testing~\cite{fan2023largelanguagemodelssoftware,Jiang_2026,yang2025surveyllmbasedautomatedprogram,sun2025sourcecodesummarizationera,10.1145/3675249.3675298}. State-of-the-art code LLMs such as DeepSeek-Coder~\cite{guo2024deepseekcoderlargelanguagemodel} and Qwen2.5-Coder~\cite{hui2024qwen25codertechnicalreport} are increasingly applied to repository-level tasks including cross-file completion, API-aware generation, and codebase question answering, which require reasoning over long, heterogeneous contexts spanning API definitions, dependency files, and historical implementations. Processing such inputs introduces three practical bottlenecks: inference cost and latency grow rapidly with context length due to the quadratic complexity of self-attention~\cite{sun2026efficientattentionmechanismslarge,hays2026attentionmechanismsneuralnetworks,wang2020linformerselfattentionlinearcomplexity}; task-critical evidence is frequently obscured by surrounding noise, a phenomenon known as ``lost in the middle''~\cite{liu-etal-2024-lost}; and finite context limits force truncation when inputs span large codebases or cross-module dependencies. These challenges are particularly acute for code, where semantic dependencies are dense and long-range, syntactic constraints are strict, and the information-to-noise ratio is considerably lower than in natural language.

Context compression \wrq{attempts to} address these bottlenecks by transforming raw context into a compact surrogate representation. Existing methods can be \wrq{categorized} into three paradigms by the representation space of their outputs. \textbf{Text-to-Text} (T2T) methods produce a shorter, discrete token sequence via token-level filtering or rewriting; representative approaches include LLMLingua~\cite{jiang2023llmlinguacompressingpromptsaccelerated}, LLMLingua-2~\cite{pan2024llmlingua2datadistillationefficient}, and LongLLMLingua~\cite{jiang2024longllmlinguaacceleratingenhancingllms}. \textbf{Text-to-Vector} (T2V) methods encode context into a compact set of continuous latent tokens; ICAE~\cite{ge2024incontextautoencodercontextcompression} distills long contexts into fixed-size memory slots, while gist-token approaches~\cite{mu2024learningcompresspromptsgist} and 500xCompressor~\cite{li2024500xcompressorgeneralizedpromptcompression} extend this paradigm to more aggressive compression ratios. \textbf{Text-to-Image} (T2I) methods render context as images consumed by a vision-language model, exploiting the higher information density of the visual modality; optical encoding has been shown to achieve up to $10\times$ token reduction while maintaining high reconstruction fidelity~\cite{wei2025deepseekocrcontextsopticalcompression,wei2026deepseekocr2visualcausal,cui2025paddleocrvlboostingmultilingualdocument}, with extensions to diverse long-context settings~\cite{feng2026agentocrreimaginingagenthistory}.

Despite this growing body of work, context compression has been studied almost exclusively on NLP tasks in method-specific settings. Whether and how these techniques transfer to code LLMs remains an open question: repository-level code imposes constraints absent in natural language, including long-range cross-file dependencies, strict syntactic structure, and a lower information-to-noise ratio that amplifies the risk of aggressive filtering. We address this gap by presenting, to the best of our knowledge, the \emph{first} systematic empirical study of context compression for repository-level code LLMs, instantiating eight representative methods across all three paradigms and studying them on code generation and completion, two of the most widely studied repository-level tasks, jointly measuring task performance and deployment efficiency across multiple model scales.

% Our study \yun{tries to answer the following}
% % is guided by 
% three research questions (RQs):
\wrq{We study} the following research questions (RQs) \wrq{and summarize the findings}:

\textbf{RQ1: How do different compression paradigms affect task performance?}
% The representation space of the compressed output is the primary determinant of downstream performance. T2V surpasses the full-context baseline by up to 28\% on completion tasks, a result we attribute to its learned bottleneck acting as a task-conditioned denoiser that amplifies task-relevant signals while suppressing boilerplate. T2I achieves completion performance within 1\% of full context at moderate compression ratios. T2T offers portable, training-free compression at mild ratios, with performance gaps widening on generation tasks.
Context compression not only preserves downstream \wrq{task performance} but can actively improve it over full-context inference. T2V surpasses the full-context baseline by up to 28.3\% \wrq{in terms of} BLEU on Python completion. We attribute \wrq{this} to the latent \wrq{vectors} acting as a learned filter that suppresses repository boilerplate and reinforces task-relevant signals. T2I tracks full-context performance closely on completion tasks and sustains competitive performance at moderate compression ratios. T2T provides training-free compression that is effective at mild ratios, though performance gaps widen on generation tasks where cross-file relational structure is critical.

\textbf{RQ2: How does compression ratio affect task performance?}
% Ratio sensitivity reflects each paradigm's information density mechanism. T2V produces near-flat performance curves from $4\times$ to $128\times$, indicating that its latent bottleneck captures task-relevant structure independent of token budget. T2I performance peaks around $4\times$ and declines monotonically with further downscaling, as uniform resolution reduction provides no semantic selection. T2T degrades most sharply: on generation tasks, performance falls to near no-context level at approximately $12\times$ compression, as perplexity-based pruning preferentially removes structurally critical yet linguistically predictable tokens.
T2V sustains its performance advantage across the full evaluated range of $4\times$ to $128\times$, with all variants exceeding full-context performance on generation at every ratio, demonstrating that effective compression does not require conservative token budgets. T2I peaks at $4\times$ on completion and degrades gradually with further downscaling, yet its generation performance remains consistently below full context regardless of ratio, as uniform rendering loses cross-file relational structure at any resolution. T2T degrades most sharply: generation performance converges to the no-context baseline at approximately $7\times$ to $12\times$, establishing a practical upper bound for perplexity-based pruning on code tasks.

\textbf{RQ3: How do compression paradigms and ratios affect throughput and resource usage?}
% Efficiency gains across all paradigms derive primarily from reduced decoding cost. T2I reduces end-to-end inference latency by up to 38\% at high compression ratios with negligible preprocessing overhead, converging to the resource cost of inference without contextual augmentation. T2V maintains a stable, ratio-independent efficiency profile, with latency and GPU memory remaining essentially constant across all evaluated ratios. T2T's efficiency frontier is governed by compressor complexity, and its most efficient configurations reach competitive latency only at ratios where task performance has already collapsed.
% All three paradigms deliver measurable efficiency gains over full-context inference, with decoding cost reduction as the primary driver across paradigms. T2I achieves a 33\% reduction in end-to-end latency at its performance-optimal $4\times$ ratio, with costs converging to near-plain levels at higher ratios. T2V maintains a stable and predictable overhead profile across all evaluated ratios, remaining well below full-context cost throughout. T2T reduces total inference latency by over 35\% relative to full-context inference at moderate compression ratios, with decoding cost and GPU memory converging toward near-plain levels at higher ratios.
All three paradigms deliver measurable efficiency gains over full-context inference, with decoding cost reduction as the primary driver across paradigms. T2I achieves a 33\% reduction in end-to-end latency at its performance-optimal $4\times$ ratio, with \wrq{resource} costs converging to the \wrq{context-free baseline} at higher \wrq{compression} ratios. T2V maintains a stable and predictable overhead profile across all evaluated ratios, remaining well below full-context cost throughout. T2T reduces total inference latency by over 35\% relative to full-context inference at moderate compression ratios, with decoding cost and GPU memory converging toward the resource cost of inference without repository context at higher ratios.

Collectively, these results demonstrate that context compression is not only viable but, under appropriate paradigm selection, performance-enhancing for repository-level code intelligence, yielding both performance improvements and substantial efficiency gains that make it a practically compelling approach for code LLM deployment.

In summary, this paper makes the following contributions:

\begin{itemize}
    \item We introduce context compression to repository-level code intelligence through the first systematic empirical study spanning three paradigms (T2T, T2V, T2I) and eight representative methods, evaluated on code generation and completion across multiple model scales.
    \item We reveal that latent-vector compression can surpass full-context performance on code tasks, establishing that compression need not be a lossy process and demonstrating paradigm-specific performance-ratio scaling behavior previously unstudied in the code domain.
    \item We provide a deployment-oriented efficiency analysis and distill actionable guidance for selecting compression strategies based on task type, compression budget, and resource constraints.
\end{itemize}

The remainder of this paper reviews related work (Section~\ref{sec:related}), presents the unified evaluation framework (Section~\ref{sec:method}), describes the experimental setup (Section~\ref{sec:setup}), reports and analyzes results (Section~\ref{sec:result}), discusses implications and threats to validity (Section~\ref{sec:discussion}), and concludes with future directions (Section~\ref{sec:conclusion}).

\section{Related Work}
\label{sec:related}

\subsection{Code Large Language Models}

Code intelligence has evolved from encoder and encoder-decoder pre-trained models such as CodeBERT~\cite{feng2020codebertpretrainedmodelprogramming}, GraphCodeBERT~\cite{guo2021graphcodebertpretrainingcoderepresentations}, CodeT5~\cite{wang2021codet5identifierawareunifiedpretrained}, and UniXcoder~\cite{guo2022unixcoderunifiedcrossmodalpretraining} to large decoder-only code LLMs including Codex~\cite{chen2021evaluatinglargelanguagemodels}, StarCoder~\cite{li2023starcodersourceyou}, CodeLlama~\cite{codellamaopenfoundation}, DeepSeek-Coder~\cite{guo2024deepseekcoderlargelanguagemodel}, and Qwen2.5-Coder~\cite{hui2024qwen25codertechnicalreport}, with progressively stronger capabilities in generation, infilling, and multi-language reasoning. General-purpose frontier models such as GPT-4~\cite{openai2024gpt4technicalreport}, Gemini~\cite{geminiteam2025geminifamilyhighlycapable}, and DeepSeek~\cite{deepseekai2025deepseekv3technicalreport} further achieve competitive coding performance through large-scale pretraining and instruction alignment.

The focus of code LLM application has shifted from function-level benchmarks to repository-level tasks including cross-file completion, project-aware bug fixing, and codebase question answering, requiring models to reason over heterogeneous, long-form contexts spanning multiple files and dependencies. Performance degrades when task-relevant signals are sparse and distributed within lengthy inputs, motivating research on context management strategies beyond model scaling. Our work investigates context compression as a deployment-oriented solution to this bottleneck, studying its applicability to repository-level code tasks.
% for the first time.

\subsection{Context Compression}

Context compression reduces prompt length while preserving task-critical information, thus lowering inference latency, memory usage, and API cost. 
We categorize existing approaches based on their distinct representation spaces:
% Existing approaches operate in three distinct representation spaces, used as the organizing framework of this paper.

\paragraph{Text-to-Text (T2T)}
T2T methods produce a shorter discrete token sequence through filtering, rewriting, or reordering without modifying the downstream model. Selective Context~\cite{li-etal-2023-compressing} removes low-information phrases using lexical self-information. LLMLingua~\cite{jiang2023llmlinguacompressingpromptsaccelerated} introduces a budget controller and small-model-guided iterative token pruning with distribution alignment. LongLLMLingua~\cite{jiang2024longllmlinguaacceleratingenhancingllms} adds a question-aware coarse-to-fine strategy with document reordering to mitigate ``lost in the middle'' effects. LLMLingua-2~\cite{pan2024llmlingua2datadistillationefficient} reformulates compression as token classification via a lightweight bidirectional encoder trained through data distillation, substantially reducing compression latency while preserving cross-domain generalization. T2T approaches are model-agnostic and produce human-readable outputs, though fidelity degrades under aggressive compression as pruning may sever long-range dependencies.

\paragraph{Text-to-Vector (T2V)}
T2V methods encode context into a compact set of continuous latent tokens, enabling the downstream model to condition on a fixed-size surrogate. Gist tokens~\cite{mu2024learningcompresspromptsgist} learn to cache demonstration semantics for reuse at test time. RMT~\cite{bulatov2022recurrentmemorytransformer} introduces segment-level recurrence with explicit memory tokens, supporting arbitrarily long inputs at bounded per-segment cost. ICAE~\cite{ge2024incontextautoencodercontextcompression} pretrains a LoRA-based encoder to compress contexts into fixed memory slots processed by a frozen LLM decoder, achieving effective $4\times$ compression. 500xCompressor~\cite{li2024500xcompressorgeneralizedpromptcompression} pushes this to extreme ratios, distilling hundreds of tokens into a single special token. R$^3$Mem~\cite{wang2025r3membridgingmemoryretention} further proposes a reversible compression architecture with hierarchical retention and high-fidelity reconstruction. T2V representations are memory-efficient and well-suited to repeated context reuse, though opaque latent encodings may attenuate exact token-level details critical for code correctness.

\paragraph{Text-to-Image (T2I)}
T2I methods render textual content as images consumed by a vision-language model, exploiting the higher information density of the visual modality. DeepSeek-OCR~\cite{wei2025deepseekocrcontextsopticalcompression} demonstrates up to $10\times$ token reduction while maintaining 97\% reconstruction fidelity at moderate compression ratios. C3~\cite{liu2025contextcascadecompressionexploring} proposes cascading models of different sizes and finds that efficiency gains may stem from latent token density rather than the image modality per se. AgentOCR~\cite{feng2026agentocrreimaginingagenthistory} extends this paradigm to agentic settings with segment optical caching, preserving over 95\% of agent performance at more than 50\% token reduction. T2I methods offer strong throughput benefits under strict token budgets, though fine-grained lexical signals such as identifiers, indentation, and delimiters may be lost under aggressive downscaling.

\noindent
These paradigms have been studied predominantly on NLP benchmarks in method-specific settings. Their effectiveness for repository-level code tasks, where long-range structural dependencies and strict correctness requirements impose notably stricter compression demands, remains unexplored. Our work addresses this gap.

\begin{figure*}[htbp]
    \centering
    \includegraphics[width=0.9\textwidth]{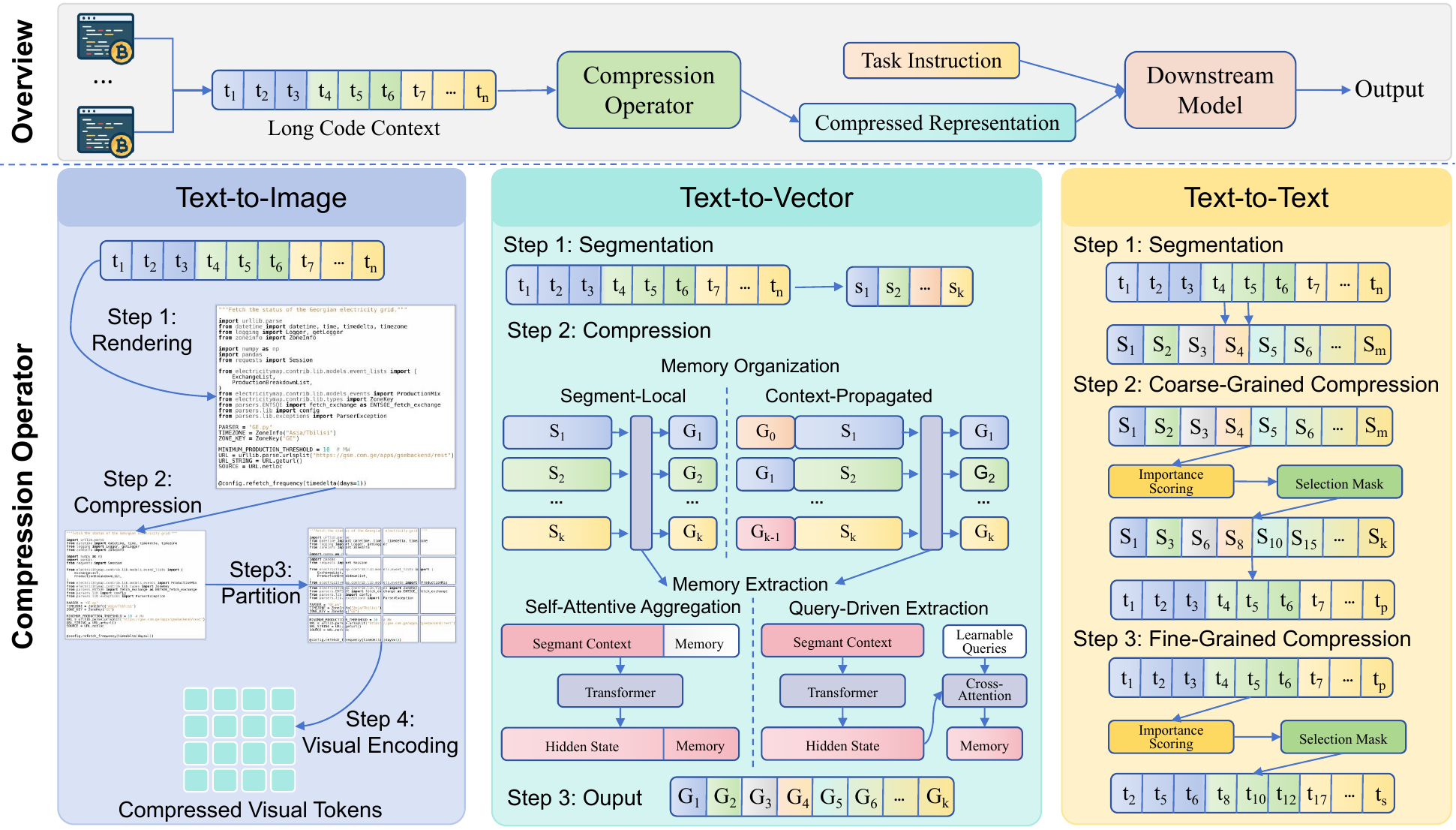}
     
    \caption{Overview of the unified context compression framework (top) and the three compression paradigms (bottom)
    % : T2I encodes context as visual tokens via rendering and patch encoding; T2V compresses segments into latent memory blocks via segment-local or context-propagated organization with self-attentive or query-driven extraction; T2T applies a two-stage coarse-to-fine importance scoring and selection over discrete tokens
    .}
   \vspace{-6pt}
    \label{fig:framework}
\end{figure*}

\section{Methodology}
\label{sec:method}

We organize the diverse compression methods studied in this paper under a common abstraction, illustrated in \autoref{fig:framework}. A compression operator transforms raw long code context into a compact surrogate representation under a resource budget, which is then consumed by a downstream model together with the task instruction. This abstraction lets us characterize methods by their output representation space and construction mechanism, independently of specific model architectures.

\subsection{Unified Problem Formulation}

% Let the input be a token sequence $\mathcal{X} = [t_1, \dots, t_n] \in \mathcal{V}^n$ and a task instruction $q \in \mathcal{Q}$. A compression operator under budget $B$ is defined as:
% \begin{equation}
% \mathcal{C}_B: \mathcal{V}^n \times \mathcal{Q} \rightarrow \mathcal{Z}, \quad \text{s.t.} \quad \text{Cost}(\mathcal{Z}) \leq B,
% \end{equation}
% where $\mathcal{Z}$ is the surrogate representation and $B$ constrains token count, memory, or latency. The downstream prediction is computed as $f(q, \mathcal{Z})$. From a rate-distortion perspective, the compression objective is:
% \begin{equation}
% \min_{\mathcal{C}_B} \mathbb{E}_{(\mathcal{X}, q, y) \sim \mathcal{D}} 
% \left[ \mathcal{L}(f(q, \mathcal{C}_B(\mathcal{X}, q)), y) \right]
% + \lambda \cdot \mathcal{R}(\mathcal{C}_B(\mathcal{X}, q)).
% \end{equation}

% Different compression paradigms primarily differ in the \emph{representation space} of $\mathcal{Z}$ and the mechanism used to construct it. As shown in \autoref{fig:framework}, this defines three paradigms: Text-to-Image (T2I), Text-to-Vector (T2V), and Text-to-Text (T2T).

Let the input be a token sequence $\mathcal{X} = [t_1, \dots, t_n] \in \mathcal{V}^n$ and a task instruction $q \in \mathcal{Q}$. A compression operator under budget $B$ maps the context to a compact surrogate:
\begin{equation}
\mathcal{C}_B: \mathcal{V}^n \rightarrow \mathcal{Z}, \quad \text{s.t.} \quad \text{Cost}(\mathcal{Z}) \leq B
\label{formula:1}
\end{equation}
where $\mathcal{Z}$ is the surrogate representation space and $B$ constrains token count, memory, or latency. The downstream prediction is $f(q, \mathcal{Z})$, with the query applied at inference time. The compression objective minimizes the output divergence from full-context inference:
\begin{equation}
\min_{\mathcal{C}_B} \mathbb{E}_{(\mathcal{X}, q) \sim \mathcal{D}}
\left[ d\!\left(f(q, \mathcal{C}_B(\mathcal{X})),\; f(q, \mathcal{X})\right) \right]
+ \lambda \cdot \mathcal{R}(\mathcal{C}_B(\mathcal{X})),
\end{equation}
where $d(\cdot,\cdot)$ measures output divergence and $\mathcal{R}(\cdot)$ is the rate penalty. Query-conditioned methods (e.g., retrieval-aware T2T) are a special case $\mathcal{C}_B(\mathcal{X};\,q)$.

% \subsection{Text-to-Image: Visual Context Encoding}
\subsection{Text-to-Image (T2I)}

The T2I paradigm, shown in the left panel of \autoref{fig:framework}, converts the code context from a token sequence into a visual representation consumed by a vision-language model (VLM). This exploits the observation that the visual modality can carry significantly more information per token than discrete text, enabling aggressive compression without modifying the downstream LLM.

% The pipeline proceeds in four stages. 
When the pipeline accepts% proceeds as follows.
the input sequence, it is first \textbf{rendered} into a high-resolution raster image via a layout-aware function:
% \begin{equation}
% \mathcal{I} = \mathcal{R}(\mathcal{X};\, \theta_{\text{layout}}),
% \end{equation}
\begin{equation}
\mathcal{I} = \text{Render}(\mathcal{X};\, \theta_{\text{layout}}),
\end{equation}
where $\theta_{\text{layout}}$ controls text density, font size, and spatial arrangement. The image is then \textbf{scaled} isotropically to a target resolution that determines the visual token budget, with the target area derived from $W \times H = N_{\text{vis}} \times p^2$ where $p$ is the patch size and $N_{\text{vis}}$ is the desired token count. The rescaled image is \textbf{partitioned} into a regular grid of $p \times p$ patches, and a vision encoder finally maps these patches into the latent representation:
\begin{equation}
\mathcal{Z} = \mathcal{E}_{\text{vis}}(\mathcal{I}),
\end{equation}
yielding $N_{\text{vis}}$ visual tokens. Crucially, this decouples the token count from the original sequence length $n$: compression ratio is controlled entirely by the target resolution, not by token filtering. The main risk is that fine-grained lexical signals such as exact identifiers and indentation may be degraded under aggressive downscaling.

% \subsection{Text-to-Vector: Latent Memory Compression}
\subsection{Text-to-Vector (T2V)}

% The T2V paradigm, shown in the middle panel of \autoref{fig:framework}, projects discrete code tokens into a continuous latent space via a learned encoder, producing a fixed-size set of memory tokens that the downstream model conditions on in place of the original sequence.
% Crucially, the downstream LLM decoder is kept entirely frozen throughout training; only the lightweight compression module (the memory extraction operators) is updated, leaving the base model's parameters and generative capabilities unchanged.
Unlike T2I, which avoids modifying the downstream model by switching to the visual modality, the T2V paradigm projects discrete code tokens into a continuous latent space via a learned encoder, producing a fixed-size set of memory tokens that the downstream model conditions on in place of the original sequence. Crucially, only the lightweight compression module is trained; the downstream LLM decoder remains entirely frozen, leaving the base model's generative capability unchanged.

The input is first divided into segments $\mathcal{X} = \{S_1, \dots, S_m\}$ for scalability. Each segment is then mapped to a compact memory block $G_i$ through two orthogonal design choices: \textbf{memory organization} and \textbf{memory extraction mechanism}.

Memory organization controls whether inter-segment context is accumulated. Under \emph{Segment-Local Memory Compression} (SLMC), each segment is encoded independently:
\begin{equation}
G_i = \text{Compress}(\mathcal{E}(S_i)),
\end{equation}
enabling parallel processing at the cost of cross-segment modeling. Under \emph{Context-Propagated Memory Compression} (CPMC), the previous memory block is carried forward:
\begin{equation}
G_i = \text{Compress}(\text{Concat}(G_{i-1},\, \mathcal{E}(S_i))),
\end{equation}
allowing progressive accumulation of context while keeping per-step memory bounded.

The extraction mechanism determines how information is distilled into $G_i$. \emph{Self-Attentive Memory Aggregation} (SAMA) lets segment representations and memory tokens interact via standard transformer attention. \emph{Query-Driven Memory Extraction} (QDME) introduces learnable queries $Q_{\text{lat}} \in \mathbb{R}^{k \times d}$ that actively retrieve information from the segment:
\begin{equation}
G_i = \text{Softmax}\!\left(\frac{(Q_{\text{lat}} W_Q)(H_{S_i} W_K)^\top}{\sqrt{d}}\right)(H_{S_i} W_V),
\end{equation}
where $k$ directly controls the compression ratio and reduces attention complexity from $\mathcal{O}(n^2)$ to $\mathcal{O}(kn)$. Combining the two memory organization strategies with the two extraction mechanisms yields four variants: SAMA-SLMC, SAMA-CPMC, QDME-SLMC, and QDME-CPMC, which we evaluate systematically in our experiments.
The primary cost of T2V relative to T2I and T2T is that the compression module must be trained before deployment, requiring a pretraining stage followed by downstream adaptation.

% \subsection{Text-to-Text: Hierarchical Saliency-Based Selection}
\subsection{Text-to-Text (T2T)}

% The T2T paradigm, shown in the right panel of \autoref{fig:framework}, performs compression entirely in the discrete token space. 
Unlike T2I and T2V, which change the representation space of the context, the T2T paradigm performs compression entirely in the discrete token space.
It retains the model-agnostic property of the original prompt while reducing its length through a \textbf{score-and-select} mechanism: each token or segment is assigned an importance score and a binary selection mask retains only the most informative units under the budget:
\begin{equation}
\omega_i = \text{Score}(u_i \mid \mathcal{X}), \qquad
\max_{M} \sum_i M_i \cdot \omega_i \quad \text{s.t.} \quad \sum_i M_i \leq B.
\end{equation}
Here, $\mathcal{X}$ and $B$ are consistent with their meanings in the \autoref{formula:1}. $u_i$ represents a token or segment within the context that needs to be compressed, and $\omega_i$ represents the importance score of it. $M_i$ is either $1$ or $0$, indicating whether or not $u_i$ is masked.

For the repository-level code tasks considered in this paper, we have applied this mechanism across two distinct stages. In the \textbf{coarse stage}, the context is segmented into structural units (e.g., code lines or functions), importance scores are computed at the segment level (e.g., via perplexity or retrieval relevance), and low-scoring segments are discarded. In the \textbf{fine stage}, token-level scores are computed within retained segments and a second selection mask produces the final compressed sequence. This \textbf{coarse-to-fine} decomposition makes the combinatorial selection problem tractable while preserving flexibility in the choice of scoring function. The compressed output remains human-readable and requires no modification to the downstream model, but fidelity can degrade at high compression ratios as aggressive pruning may sever long-range cross-file dependencies.

\section{Experimental Setup}
\label{sec:setup}

We implement all compression methods within the unified pipeline described in \autoref{sec:method}. Each paradigm uses its corresponding downstream model (Qwen2.5-VL for T2I, Qwen2.5-Coder for T2V and T2T), with the task instruction $q$ kept identical across all settings so that observed differences arise solely from the compression operator $\mathcal{C}_B$. For each paradigm, we sweep a range of compression ratios to support the cross-paradigm comparison in RQ1--RQ3.

\subsection{Subject Models}

We select models from the Qwen2.5 family~\cite{qwen2025qwen25technicalreport} at two scales, covering both code-specialized and multimodal variants.

\begin{itemize}
    \item \textbf{Qwen2.5-Coder (QC)} is a code-specialized language model family with strong capabilities in code completion, infilling, and long-context reasoning. We use QC at 3B and 7B scales to instantiate the T2V and T2T paradigms.
    \item \textbf{Qwen2.5-VL (QV)} is a vision-language model family supporting joint visual and textual understanding. We use QV at 3B and 7B scales to instantiate the T2I paradigm.
\end{itemize}

\subsection{Tasks and Benchmark}

We evaluate on two generation-oriented code intelligence tasks using \textbf{ComplexCodeEval}~\cite{Feng_2024}, a benchmark designed to reflect realistic repository-level coding scenarios with structured inputs including function signatures, docstrings, and API usages.

\textbf{Code generation} requires synthesizing a complete function body from a natural language specification (signature and docstring), testing whether compressed representations preserve sufficient semantic content for correct program synthesis.

\textbf{Code completion} requires predicting the missing portion of a partially observed code snippet given its surrounding context, evaluating how well compressed representations retain local and cross-file contextual cues under a token budget.

Following the original benchmark setup, we randomly sample 100 instances from each of the Java and Python subsets as a held-out evaluation set of 200 instances. For T2V, we additionally use the ComplexCodeEval training split (10k+ samples) for downstream adaptation of the compression module,
% (the downstream LLM is never updated),
with the 200 evaluation instances withheld from all training stages.

\subsection{Compared Methods}

We compare methods from all three paradigms against two non-compression baselines:

\begin{itemize}
    \item \textbf{Full Context}: the uncompressed repository context is fed directly to the downstream model, serving as the performance upper bound.
    \item \textbf{No Context}: only the task instruction is provided without any context, representing the lower bound.
\end{itemize}

For each paradigm, we evaluate the following representative methods:

\textbf{T2I.} We implement a rendering-based pipeline that converts the code context into a compact visual representation consumed by Qwen2.5-VL (see \autoref{sec:impl-t2i}).

\textbf{T2V.} We evaluate all four combinations of memory organization and extraction mechanism defined in \autoref{sec:method}: \emph{SAMA-SLMC} (T2V-SS), \emph{SAMA-CPMC} (T2V-SC), \emph{QDME-SLMC} (T2V-QS), and \emph{QDME-CPMC} (T2V-QC).
% , where SAMA = Self-Attentive Memory Aggregation, QDME = Query-Driven Memory Extraction, SLMC = Segment-Local Memory Compression, and CPMC = Context-Propagated Memory Compression.

\textbf{T2T.} We evaluate three methods from the LLMLingua family~\cite{jiang2023llmlinguacompressingpromptsaccelerated,jiang2024longllmlinguaacceleratingenhancingllms,pan2024llmlingua2datadistillationefficient}: \emph{LLMLingua} removes low-information tokens based on small-model perplexity; \emph{LongLLMLingua} extends this with query-aware scoring and reordering strategies to mitigate ``lost-in-the-middle'' effects; \emph{LLMLingua-2} reformulates compression as token classification using a lightweight bidirectional encoder, reducing latency while maintaining cross-domain generalization.

\subsection{Implementation Details}

\begin{table*}[ht]

\centering

\caption{Comparison of compression paradigms at $\sim$4$\times$ budget on ComplexCodeEval. QC\,=\,Qwen2.5-Coder, QV\,=\,Qwen2.5-VL; T2V variants SS/SC/QS/QC denote SAMA-SLMC/CPMC and QDME-SLMC/CPMC; all metrics higher is better. Bold indicates the optimal result; underline indicates the second-best result (all subsequent tables adhere to this setting).}
\vspace{-6pt}
\renewcommand{\arraystretch}{0.9}
\label{tab:unified_results}

% \small

\begin{tabular}{ccc|ccc|ccc||cc|cc}

\toprule

\multirow{3}{*}{\textbf{Model}} & \multirow{3}{*}{\textbf{Method}} & \multirow{3}{*}{\textbf{Ratio}} 

& \multicolumn{6}{c||}{\textbf{Code Completion}} 

& \multicolumn{4}{c}{\textbf{Code Generation}} \\

\cmidrule(lr){4-9} \cmidrule(lr){10-13}

& & 

& \multicolumn{3}{c}{Python} 

& \multicolumn{3}{c||}{Java} 

& \multicolumn{2}{c}{Python} 

& \multicolumn{2}{c}{Java} \\

\cmidrule(lr){4-6} \cmidrule(lr){7-9} \cmidrule(lr){10-11} \cmidrule(lr){12-13}

& & 

& BLEU & ES & EM 

& BLEU & ES & EM 

& BLEU & ES 

& BLEU & ES \\

\midrule

\multirow{3}{*}{QV-3B}

& plain   & --   & 10.32 & 34.03 & 8.00  & 14.58 & 36.66 & 12.00 & 4.07 & 16.72 & 4.98 & 16.00 \\

& context & --   & \textbf{18.03} & \textbf{40.56} & \textbf{17.00} & \underline{17.77} & \textbf{43.52} & \underline{15.00} & \textbf{8.36} & \textbf{24.71} & \textbf{12.01} & \textbf{29.34} \\

& T2I     & 4.0$\times$ & \underline{14.57} & \underline{39.93} & \underline{12.00} & \textbf{19.27} & \underline{41.27} & \textbf{17.00} & \underline{4.51} & \underline{19.83} & \underline{3.40} & \underline{15.77} \\

\midrule

\multirow{3}{*}{QV-7B}

& plain   & --   & 20.29 & 43.69 & 19.00 & 16.52 & 38.90 & 15.00 & 4.18 & 16.44 & 5.72 & 18.94 \\

& context & --   & \textbf{24.91} & \textbf{48.98} & \underline{25.00} & \textbf{20.47} & \textbf{45.02} & \textbf{19.00} & \textbf{9.19} & \textbf{25.46} & \textbf{14.49} & \textbf{31.29} \\

& T2I     & 4.0$\times$ & \underline{24.11} & \underline{48.52} & \textbf{26.00} & \underline{19.29} & \underline{43.91} & \underline{18.00} & \underline{4.52} & \underline{17.29} & \underline{7.91} & \underline{21.73} \\

\midrule

\multirow{9}{*}{QC-3B}

& plain   & --   & 27.59 & 52.77 & 28.00 & 25.60 & 49.29 & 23.00 & 5.32 & 18.90 & 7.19 & 21.18 \\

& context & --   & 31.83 & 60.19 & 34.00 & \textbf{31.70} & \textbf{59.82} & \textbf{34.00} & 10.27 & 28.75 & \textbf{16.58} & 35.45 \\

& T2V-SS  & 4.0$\times$ & 34.48 & 59.93 & \underline{38.00} & 29.37 & 53.51 & 27.00 & 10.99 & 33.16 & 12.49 & 35.54 \\

& T2V-SC  & 4.0$\times$ & 33.04 & 58.32 & 35.00 & 27.91 & 52.06 & 25.00 & 11.94 & \textbf{34.16} & \underline{15.47} & \textbf{37.79} \\

& T2V-QS  & 4.0$\times$ & \textbf{36.76} & \textbf{62.59} & \textbf{41.00} & 29.32 & 54.25 & 27.00 & \underline{12.09} & 33.18 & 15.42 & \underline{37.64} \\

& T2V-QC  & 4.0$\times$ & \underline{35.06} & \underline{60.80} & 37.00 & \underline{31.11} & \underline{54.69} & \underline{29.00} & \textbf{12.43} & \underline{33.41} & 14.84 & 37.52 \\

& T2T-LL  & 3.7$\times$ & 25.69 & 51.93 & 27.00 & 27.89 & 52.33 & 27.00 & 7.17  & 23.08 & 8.86  & 23.99 \\

& T2T-LL2 & 3.7$\times$ & 25.99 & 54.67 & 26.00 & 29.44 & 54.12 & 27.00 & 6.21  & 21.79 & 9.75  & 26.93 \\

& T2T-LLL & 4.6$\times$ & 25.78 & 51.64 & 26.00 & 29.00 & 53.56 & 26.00 & 6.99  & 22.07 & 8.05  & 22.81 \\

\midrule

\multirow{9}{*}{QC-7B}

& plain   & --   & 29.74 & 54.97 & 27.00 & 28.95 & 50.45 & 28.00 & 6.14 & 20.27 & 7.60 & 21.35 \\

& context & --   & 32.21 & 56.64 & 33.00 & 32.60 & 55.90 & \underline{34.00} & 10.49 & 28.25 & \textbf{17.92} & 36.09 \\

& T2V-SS  & 4.0$\times$ & \textbf{41.34} & \textbf{67.51} & \underline{42.00} & 28.87 & 53.78 & 27.00 & \underline{12.83} & \underline{34.19} & 15.68 & \underline{38.46} \\

& T2V-SC  & 4.0$\times$ & 38.16 & 62.78 & 41.00 & 31.03 & 54.75 & 29.00 & 12.41 & 33.32 & 15.79 & 38.33 \\

& T2V-QS  & 4.0$\times$ & \underline{40.03} & \underline{65.77} & \textbf{44.00} & \underline{33.44} & \textbf{60.29} & 33.00 & \textbf{13.58} & \textbf{34.27} & \underline{17.23} & \textbf{39.86} \\

& T2V-QC  & 4.0$\times$ & 37.24 & 63.75 & 40.00 & \textbf{35.12} & \underline{60.08} & \textbf{35.00} & 12.19 & 32.90 & 13.26 & 35.25 \\

& T2T-LL  & 3.7$\times$ & 29.69 & 55.11 & 29.00 & 29.01 & 52.14 & 28.00 & 6.89  & 23.08 & 9.71  & 25.09 \\

& T2T-LL2 & 3.7$\times$ & 30.32 & 54.22 & 29.00 & 28.14 & 51.80 & 27.00 & 7.54  & 24.01 & 10.55 & 26.21 \\

& T2T-LLL & 4.6$\times$ & 30.07 & 56.92 & 30.00 & 31.16 & 55.86 & 30.00 & 7.30  & 22.88 & 10.22 & 25.89 \\

\bottomrule

\end{tabular}
\vspace{-6pt}
\renewcommand{\arraystretch}{1}

\end{table*}

\subsubsection{Text-to-Image (T2I)}
\label{sec:impl-t2i}

% Our T2I pipeline processes source code through three stages before feeding the result into Qwen2.5-VL.

% \emph{(1) Master rendering.} The source code is rasterized into a high-resolution master image with automatic line wrapping and padding margins. Rendering uses high-contrast black/white RGB to preserve sharp character boundaries for subsequent downscaling.

% \emph{(2) Token-constrained scaling.} Given a target visual token budget $N$, we compute target image dimensions from $W \times H = N \times 28^2$ (one visual token corresponds to a $28 \times 28$ patch in Qwen2.5-VL), with both $W$ and $H$ aligned to multiples of $28$ to avoid redundant padding during patch partition.

% \emph{(3) High-quality resampling.} We apply LANCZOS resampling for downscaling, which suppresses aliasing and preserves text edges particularly at high compression ratios, improving visual readability for the downstream VLM.

% Compression ratio is precisely controlled by the target visual token budget $N$: setting $N = \lfloor n / r \rfloor$ where $n$ is the original context token count yields compression ratio $r$. We evaluate $r \in \{2, 4, 8, 16, 32, 64, 128\}$.

We implement the T2I pipeline described in \autoref{sec:method} on top of Qwen2.5-VL, where one visual token corresponds to a $28 \times 28$ patch. Source code is rasterized in high-contrast black/white RGB; images are isotropically scaled with $W$ and $H$ aligned to multiples of 28 and downsampled with LANCZOS resampling~\cite{lanczos1964evaluation} to preserve text edges at high compression ratios. Compression ratio is controlled by the target visual token budget $N = \lfloor n / r \rfloor$, where $n$ is the original token count. We evaluate $r \in \{2, 4, 8, 16, 32, 64, 128\}$.

\subsubsection{Text-to-Vector (T2V)}

Following the standard two-stage protocol in prior T2V work~\cite{ge2024incontextautoencodercontextcompression,li2024500xcompressorgeneralizedpromptcompression}, we train only the memory extraction operators while keeping the downstream LLM decoder frozen throughout. The two-stage design is necessary because pretraining on a text-recovery objective alone does not yield memory tokens that optimally steer the frozen decoder toward task-specific outputs; downstream adaptation closes this gap without modifying the base model.

\textbf{Pretraining} uses StarCoderData~\cite{li2023starcodersourceyou} (100K Python + 100K Java). The encoder compresses the input into memory tokens, which the frozen LLM decodes; only the extraction module receives gradient updates. We train for 3 epochs with BF16 mixed-precision, learning rate $1\times10^{-4}$, cosine decay with 100 warmup steps, effective batch size 256, and sequence length 16,384 with gradient checkpointing.

\textbf{Downstream fine-tuning} initializes from the pretrained checkpoint and adapts on the ComplexCodeEval training split for 3 epochs under the same learning rate schedule, using 8-step gradient accumulation to handle long-context inputs. The base Qwen2.5-Coder model is never updated at any stage. Compression ratio is controlled by $k$, the number of memory tokens per segment (query count in QDME, memory slot count in SAMA); we evaluate $r \in \{4, 8, 16, 32, 64, 128\}$ by adjusting $k$ per segment size.

\subsubsection{Text-to-Text (T2T)}

We apply LLMLingua, LongLLMLingua, and LLMLingua-2 with their default configurations. All three methods operate on the discrete token sequence and require no task-specific training on the evaluation data.

Unlike T2I and T2V, T2T methods accept a target compression ratio as a soft threshold and prune tokens accordingly; the actual achieved ratio may deviate from the target, as it depends on the input-specific token distribution and each method's scoring function. We set target ratios of $\{2, 4, 8, 16, 32\}$ for all three methods.

\subsubsection{Inference and Hardware}

All methods use greedy decoding (temperature $= 0$, top-$p = 1$). Standard inference runs with \texttt{vLLM}. For RQ3 efficiency profiling, we switch to native Hugging Face Transformers inference to minimize framework-induced runtime variance; throughput is reported as the inverse of end-to-end latency under identical hardware and batch settings. All experiments are conducted on a server with 8$\times$A100 80GB GPUs; each individual training or inference run uses a single GPU.

\subsection{Evaluation Metrics}

We follow the official ComplexCodeEval evaluation protocol~\cite{Feng_2024} and report task-specific automatic metrics.

\textbf{Code generation.} We report \textbf{BLEU} and \textbf{Edit Similarity} to measure n-gram overlap and string-level structural similarity between generated code and references.

\textbf{Code completion.} We additionally report \textbf{Exact Match (EM)}, which measures whether the predicted completion exactly matches the reference.

\section{Result}
\label{sec:result}

\begin{figure*}[htbp]
    \centering
    \includegraphics[width=0.9\textwidth]{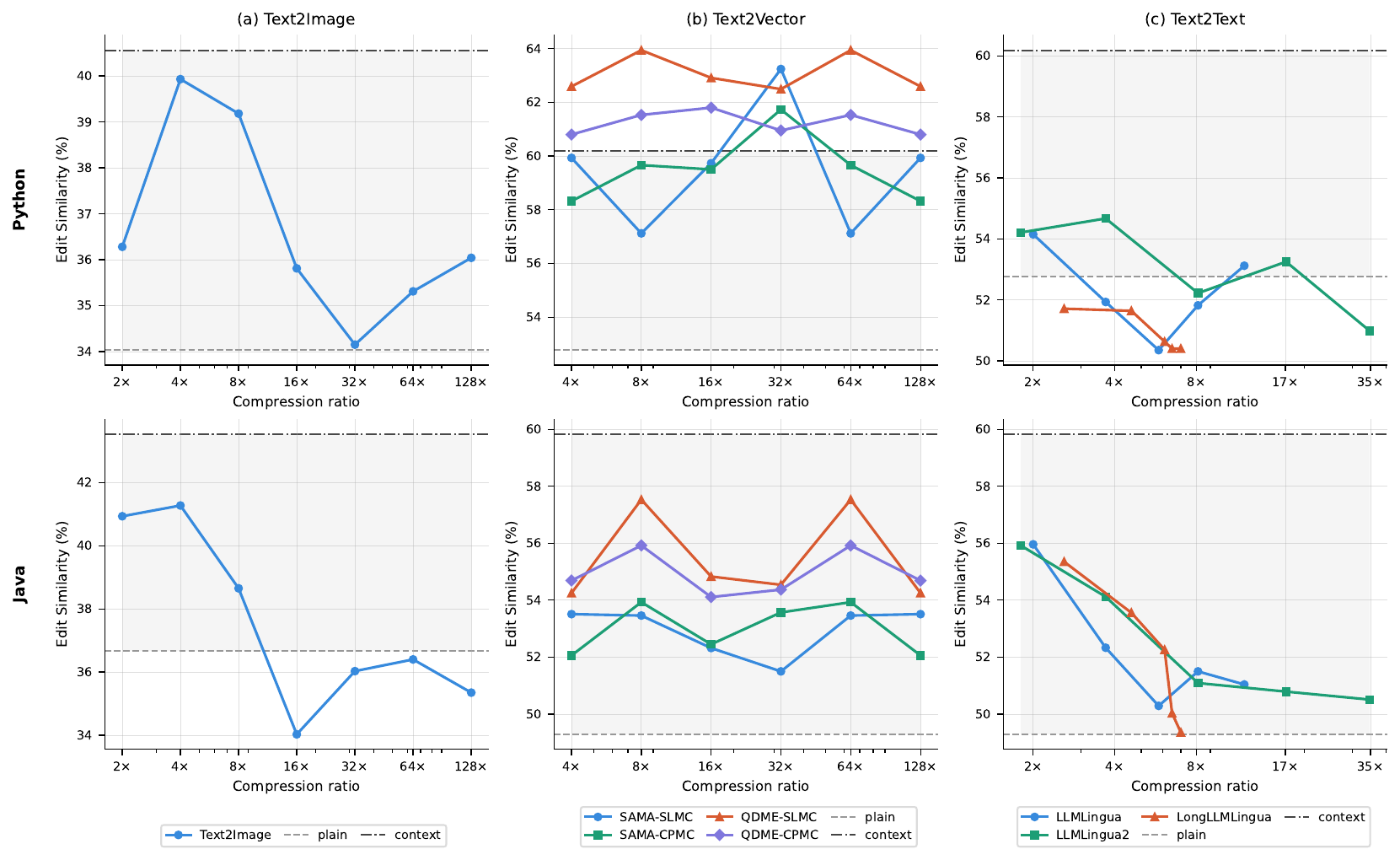}
    \vspace{-6pt}
    \caption{Edit Similarity (\%) on code completion across compression ratios for QC-3B (T2V, T2T) and QV-3B (T2I). Dashed lines indicate no-context (\textit{plain}) and full-context (\textit{context}) baselines; higher is better.}
    \label{fig:rq2_completion_es}
\end{figure*}

\subsection{RQ1: How do different compression paradigms affect task performance?}

\autoref{tab:unified_results} shows that at $\sim$4$\times$ compression, paradigm choice is the primary factor determining whether compression improves or degrades downstream performance.

\textbf{T2V: the latent bottleneck filters repository noise.}
On QC-7B Python, all four T2V variants exceed the full-context baseline on both completion (T2V-SS: 41.34/67.51 BLEU/ES vs.\ context 32.21/56.64, +28.3\% BLEU) and generation (T2V-QS: 13.58/34.27 vs.\ context 10.49/28.25, +29.5\% BLEU). At QC-3B, QDME variants exceed context on Python completion (T2V-QS: 36.76 vs.\ 31.83) but not on Java, suggesting sufficient model capacity is required to realize the supra-context effect. Repository context is informationally sparse: raw multi-file sequences are dominated by boilerplate, redundant declarations, and import chains that dilute task-specific signal. The latent bottleneck is trained to reconstruct task-relevant content under a fixed memory budget, effectively acting as a learned filter rather than a truncator. Variant selection is language-sensitive: segment-local encoding dominates on QC-7B Python completion (SS $>$ QS $>$ SC $>$ QC), while context-propagated organization is preferred on Java (QC $>$ QS $>$ SC $>$ SS), consistent with Python's function-level modularity versus Java's cross-class dependencies.

\begin{tcolorbox}[colback=gray!10, colframe=gray!50, boxrule=0.5pt, left=4pt, right=4pt, top=3pt, bottom=3pt]
\textbf{Finding 1.} T2V surpasses full-context performance on code tasks, achieving up to +28.3\% BLEU on Python completion at 7B scale. The magnitude of improvement is language- and scale-dependent: gains are consistent across all variants on Python at 7B but weaker or absent on Java at 3B scale.
\end{tcolorbox}

\textbf{T2I: uniform rendering causes non-selective information loss.}
T2I closely tracks full context on completion (QV-7B Python: 24.11/48.52 vs.\ 24.91/48.98; QV-3B Java: 19.27/41.27, marginally above context 17.77/43.52), but degrades sharply on generation (QV-7B Python: 4.52/17.29 vs.\ context 9.19/25.46, $-$50.8\% BLEU; QV-3B Java: 3.40/15.77, below the no-context baseline 4.98/16.00). This divergence stems from how rendering operates: every token is mapped to an equal image area with no mechanism to emphasize task-relevant regions, so uniform downscaling degrades function signatures, API chains, and cross-file imports at the same rate as surrounding boilerplate. Completion is tolerant because local syntactic cues survive moderate resolution reduction; generation requires intact cross-file relational structure that non-selective fidelity loss cannot preserve.

\begin{tcolorbox}[colback=gray!10, colframe=gray!50, boxrule=0.5pt, left=4pt, right=4pt, top=3pt, bottom=3pt]
\textbf{Finding 2.} T2I performs competitively on completion but degrades sharply on generation, falling up to $-$50.8\% BLEU below full context on Python at 7B scale. This task asymmetry is consistent across model sizes and languages.
\end{tcolorbox}

\textbf{T2T: perplexity-based scoring misaligns with code structure.}
On QC-3B Python completion, T2T-LL (25.69 BLEU) and T2T-LLL (25.78) fall \emph{below} the no-context baseline (27.59), while the same methods remain above plain on Java (T2T-LL: 27.89; T2T-LLL: 29.00). This language asymmetry reflects a structural property of Python: indentation markers are syntactically load-bearing but locally predictable, making them early pruning targets. Their removal corrupts scope boundaries and control flow, causing the pruned context to actively mislead the model. Java's explicit brace delimiters are more robust to token removal, which explains why T2T completion remains competitive on Java. On generation, the collapse is consistent across both languages, as import declarations and function signatures are regular patterns assigned low perplexity and thus removed before higher-ratio compression engages; once severed, cross-file dependencies cannot be recovered (T2T-LL QC-7B Python generation: 6.89/23.08 vs.\ plain 6.14/20.27).

\begin{tcolorbox}[colback=gray!10, colframe=gray!50, boxrule=0.5pt, left=4pt, right=4pt, top=3pt, bottom=3pt]
\textbf{Finding 3.} Perplexity-based pruning degrades code task performance in a language-asymmetric manner: T2T falls below the no-context baseline on Python completion at 3B scale while remaining above it on Java. On generation, performance reaches no-context level at moderate compression ratios across both languages.
\end{tcolorbox}

\subsection{RQ2: How does compression ratio affect performance?}

\begin{figure*}[htbp]
    \centering
    \includegraphics[width=0.9\textwidth]{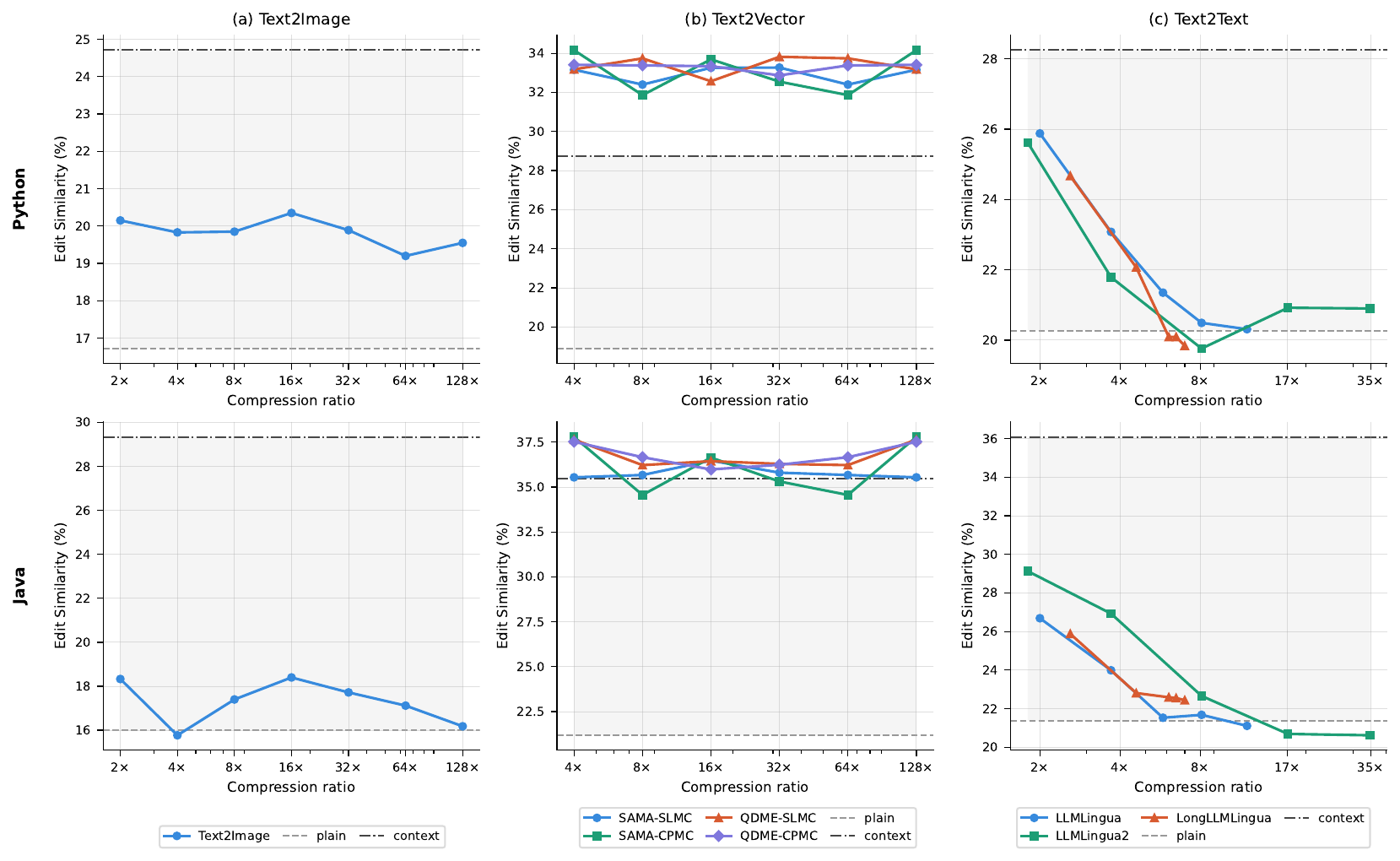}
    \vspace{-6pt}
    \caption{Edit Similarity (\%) on code generation across compression ratios for QC-3B (T2V, T2T) and QV-3B (T2I). Dashed lines indicate no-context (\textit{plain}) and full-context (\textit{context}) baselines; higher is better.}
    \vspace{-6pt}
    \label{fig:rq2_generation_es}
\end{figure*}

Figures~\ref{fig:rq2_completion_es} and~\ref{fig:rq2_generation_es} report ES across compression ratios from 2$\times$ to 128$\times$. Each paradigm's ratio sensitivity is a direct extension of the compression mechanism identified in RQ1.

\textbf{T2V performance is insensitive to compression ratio.}
All four T2V variants produce near-flat ES curves from 4$\times$ to 128$\times$ on both tasks. On Python completion, QDME-SLMC ranges between 62.49\% and 63.94\% (context: 60.19\%); on generation, SAMA-SLMC holds between 32.40\% and 33.27\% (context: 28.75\%). All four variants exceed full context on generation at every ratio across both languages, a stronger result than on completion, where Java variants remain below context throughout. The fixed memory bottleneck extracts task-relevant structure regardless of input length; increasing ratio reallocates the same latent representation rather than discarding additional information.

\textbf{T2I completion and generation respond differently to ratio.}
On completion, T2I peaks at 4$\times$ (Python: 39.93\%; Java: 41.27\%), degrades beyond 8$\times$, and partially recovers at high ratios (Python: 34.15\% at 32$\times$ to 36.04\% at 128$\times$; Java: from 34.03\% at 16$\times$ to 35.35\% at 128$\times$, below plain 36.66\% at most ratios). On generation, ES is nearly flat across all ratios (Python: 19.20\%--20.35\%; Java: 15.77\%--18.40\%), well below context at every point. Local syntactic cues sufficient for completion survive moderate resolution reduction, so ratio governs how much of this signal is preserved; cross-file relational structure required for generation is lost at any resolution, making the generation gap ratio-independent.

\textbf{T2T generation utility collapses at a method-specific critical ratio.}
On completion, all three methods degrade gradually and remain above plain across most of the evaluated range (LLMLingua2 Python: 54.67\% at 3.7$\times$ to 50.98\% at 34.8$\times$, vs.\ plain 52.77\%). On generation, degradation is substantially faster: LLMLingua Python ES reaches 20.31\% at 12$\times$, equalling the no-context baseline (20.27\%); LLMLingua2 falls below plain at 8.1$\times$ (19.76\%); LongLLMLingua crosses the threshold at 7$\times$ on both Python (19.84\%) and Java (22.44\% vs.\ plain 21.35\%). Python reaches these critical ratios consistently earlier than Java, in line with its higher structural sensitivity to perplexity-based pruning (Finding 3).

\begin{tcolorbox}[colback=gray!10, colframe=gray!50, boxrule=0.5pt, left=4pt, right=4pt, top=3pt, bottom=3pt]
\textbf{Finding 4.} T2V performance is stable across $4\times$--$128\times$, with the supra-context generation advantage maintained at all evaluated ratios. T2I completion peaks at $4\times$ and degrades with further compression, while generation remains nearly flat and consistently below full context regardless of ratio. T2T generation reaches no-context level at $7\times$--$12\times$, with Python reaching this threshold at lower ratios than Java.
\end{tcolorbox}

\subsection{RQ3: How do compression methods/ratios affect throughput and resource usage?}
\begin{table}[ht]

\centering

\caption{T2I latency breakdown (s/sample) and peak GPU memory under increasing compression ratios; lower is better for all columns.}
\vspace{-6pt}
\renewcommand{\arraystretch}{0.9}
\label{tab:t2i_eff}

\small

\resizebox{\columnwidth}{!}{%

\begin{tabular}{lccccc}

\toprule

\textbf{Method} & \textbf{Ratio} & \textbf{Comp.} & \textbf{Decod.} & \textbf{Total} & \textbf{GPU (GB)} \\

\midrule

Plain   & --    & 0.00 & 4.17 & 4.17 & 7.09 \\

Context & --    & 0.00 & 8.95 & 8.95 & 10.78 \\

\midrule

\multirow{7}{*}{\textit{Text-to-Image}} & 2.0$\times$   & 0.42 & 6.78 & 7.20 & 8.54 \\

             & 4.0$\times$   & 0.36 & 5.67 & 6.02 & 7.79 \\

             & 8.0$\times$   & 0.34 & 5.50 & 5.83 & 7.39 \\

             & 16.0$\times$  & 0.31 & 5.18 & 5.50 & 7.22 \\

             & 32.0$\times$  & \underline{0.30} & 4.83 & 5.13 & 7.14 \\

             & 64.0$\times$  & \underline{0.30} & \underline{4.47} & \underline{4.76} & \underline{7.11} \\

             & 128.0$\times$ & \textbf{0.29} & \textbf{4.19} & \textbf{4.48} & \textbf{7.10} \\

\bottomrule

\end{tabular}%

}
\renewcommand{\arraystretch}{1}
\end{table}
\begin{table}[ht]
\centering
\caption{T2V latency breakdown (s/sample) and peak GPU memory across all four variants and compression ratios; lower is better for all columns.}
\vspace{-6pt}
\renewcommand{\arraystretch}{0.9}
\label{tab:t2v_eff_full}
\small
\resizebox{\columnwidth}{!}{%
\begin{tabular}{lccccc}
\toprule
\textbf{Variant} & \textbf{Ratio} & \textbf{Comp.} & \textbf{Decod.} & \textbf{Total} & \textbf{GPU (GB)} \\
\midrule
Plain   & -- & 0.00 & 4.21 & 4.21 & 5.85 \\
Context & -- & 0.00 & 9.13 & 9.13 & 10.92 \\
\midrule
\multirow{6}{*}{\textit{SAMA-SLMC}}
% \textit{SAMA-SLMC}
& 4.0$\times$   & \underline{0.22} & 6.29 & 6.51 & 6.57 \\
& 8.0$\times$   & 0.23 & 6.03 & 6.26 & 6.51 \\
& 16.0$\times$  & \underline{0.22} & 6.08 & 6.30 & 6.48 \\
& 32.0$\times$  & \textbf{0.21} & \underline{5.54} & \underline{5.75} & \underline{6.46} \\
& 64.0$\times$  & \underline{0.22} & 5.71 & 5.93 & \textbf{6.45} \\
& 128.0$\times$ & 0.25 & \textbf{5.46} & \textbf{5.71} & \textbf{6.45} \\
\midrule
\multirow{6}{*}{\textit{SAMA-CPMC}}
% \textit{SAMA-CPMC} 
& 4.0$\times$   & \textbf{0.24} & 6.38 & 6.63 & 6.68 \\
& 8.0$\times$   & \underline{0.25} & 6.28 & 6.53 & 6.53 \\
& 16.0$\times$  & \underline{0.25} & 6.17 & 6.41 & 6.47 \\
& 32.0$\times$  & \underline{0.25} & \underline{6.09} & \underline{6.34} & 6.44 \\
& 64.0$\times$  & \underline{0.25} & 6.28 & 6.53 & \underline{6.43} \\
& 128.0$\times$ & \textbf{0.24} & \textbf{5.82} & \textbf{6.07} & \textbf{6.42} \\
\midrule
\multirow{6}{*}{\textit{QDME-SLMC}}
% \textit{QDME-SLMC} 
& 4.0$\times$   & \textbf{0.22} & 6.21 & 6.42 & \underline{6.52} \\
& 8.0$\times$   & \textbf{0.22} & 6.00 & 6.22 & \textbf{6.51} \\
& 16.0$\times$  & \underline{0.23} & 6.05 & 6.28 & \textbf{6.51} \\
& 32.0$\times$  & \textbf{0.22} & \underline{5.98} & \underline{6.20} & \textbf{6.51} \\
& 64.0$\times$  & \textbf{0.22} & 6.05 & 6.26 & \textbf{6.51} \\
& 128.0$\times$ & \textbf{0.22} & \textbf{5.82} & \textbf{6.04} & \textbf{6.51} \\
\midrule
\multirow{6}{*}{\textit{QDME-CPMC}}
% \textit{QDME-CPMC} 
& 4.0$\times$   & \textbf{0.25} & 6.49 & 6.74 & 6.60 \\
& 8.0$\times$   & \textbf{0.25} & \underline{6.21} & \underline{6.46} & 6.54 \\
& 16.0$\times$  & 0.27 & 6.25 & 6.52 & 6.51 \\
& 32.0$\times$  & \textbf{0.25} & \underline{6.21} & \underline{6.46} & \underline{6.49} \\
& 64.0$\times$  & \underline{0.26} & 6.40 & 6.66 & \textbf{6.48} \\
& 128.0$\times$ & \underline{0.26} & \textbf{6.19} & \textbf{6.45} & \textbf{6.48} \\
\bottomrule
\end{tabular}%
}
\renewcommand{\arraystretch}{1}
\end{table}
\begin{table}[ht]
\centering
\caption{T2T latency breakdown (s/sample) and peak GPU memory across LLMLingua-family methods and compression ratios; lower is better for all columns.}
\vspace{-6pt}
\renewcommand{\arraystretch}{0.9}
\label{tab:t2t_eff}
\small
\resizebox{\columnwidth}{!}{%
\begin{tabular}{lccccc}
\toprule
\textbf{Method} & \textbf{Ratio} & \textbf{Comp.} & \textbf{Decod.} & \textbf{Total} & \textbf{GPU (GB)} \\
\midrule
Plain   & -- & 0.00 & 4.21 & 4.21 & 5.85 \\
Context & -- & 0.00 & 9.13 & 9.13 & 10.92 \\
\midrule
\multirow{5}{*}{\textit{LLMLingua}} & 2.0$\times$ & \underline{0.64} & 6.15 & 6.80 & 6.54 \\
                   & 3.7$\times$ & 0.66 & 5.17 & 5.83 & 6.16 \\
                   & 5.8$\times$ & \textbf{0.63} & \underline{4.67} & \underline{5.30} & 6.01 \\
                   & 8.1$\times$ & \textbf{0.63} & 4.75 & 5.38 & \underline{5.94} \\
                   & 12.0$\times$ & \textbf{0.63} & \textbf{4.45} & \textbf{5.08} & \textbf{5.90} \\
\midrule
\multirow{5}{*}{\textit{LLMLingua2}} & 1.8$\times$ & \textbf{0.22} & 7.31 & 7.52 & 6.69 \\
                    & 3.7$\times$ & \textbf{0.22} & 7.05 & 7.27 & 6.26 \\
                    & 8.1$\times$ & \textbf{0.22} & 5.33 & 5.54 & 6.03 \\
                    & 17.1$\times$ & \textbf{0.22} & \underline{4.77} & \underline{4.98} & \underline{5.93} \\
                    & 34.8$\times$ & \textbf{0.22} & \textbf{4.36} & \textbf{4.58} & \textbf{5.88} \\
\midrule
\multirow{5}{*}{\textit{LongLLMLingua}} & 2.6$\times$ & 1.36 & 5.43 & 6.80 & 6.75 \\
                       & 4.6$\times$ & 1.38 & 4.78 & 6.15 & 6.26 \\
                       & 6.1$\times$ & 1.30 & 4.63 & 5.93 & 6.05 \\
                       & 6.5$\times$ & \underline{1.28} & \textbf{4.29} & \underline{5.57} & \underline{5.91} \\
                       & 7.0$\times$ & \textbf{1.24} & \underline{4.30} & \textbf{5.54} & \textbf{5.87} \\
\bottomrule
\end{tabular}%
}
\renewcommand{\arraystretch}{1}
\vspace{-1.0em}
\end{table}

Tables~\ref{tab:t2i_eff}--\ref{tab:t2t_eff} report latency and GPU memory for QV-3B (T2I) and QC-3B (T2V, T2T). All methods remain well below full-context overhead across all ratios, with decoding dominating the total costs.

\textbf{T2I achieves the most pronounced efficiency improvement with ratio.}
Compression and decoding latency decrease monotonically: compression falls from 0.42 s at 2$\times$ to 0.29 s at 128$\times$; decoding from 6.78 s to 4.19 s, converging to plain (4.17 s). GPU memory follows the same trend, from 8.54 GB to 7.10 GB, nearly matching plain (7.09 GB). At 4$\times$, the performance peak for completion (Finding 4), total latency is already 33\% below full context (6.02 s vs.\ 8.95 s); at 128$\times$, the overhead above plain reduces to 0.31 s and 0.01 GB.

\textbf{T2V maintains stable compression cost while decoding latency decreases gradually.}
Compression overhead is small and ratio-independent across all variants (0.21--0.27 s). Decoding latency decreases with ratio (SAMA-SLMC: 6.29 s at 4$\times$ to 5.46 s at 128$\times$; QDME-SLMC: 6.21 s to 5.82 s), with a modest accompanying decrease in GPU memory (SAMA-SLMC: 6.57 GB to 6.45 GB). All variants remain well below full context at every ratio, and the gap to plain narrows to approximately 1.5 s and 0.6 GB at high ratios. CPMC variants incur slightly higher compression overhead than SLMC counterparts (0.24--0.27 s vs.\ 0.21--0.23 s), consistent with the additional cross-segment computation described in \autoref{sec:method}.

\textbf{T2T compression cost is method-dependent and ratio-stable.}
Compression overhead varies substantially across methods but is stable within each: LLMLingua2 costs 0.22 s (comparable to T2V), LLMLingua 0.63--0.66 s, and LongLLMLingua 1.24--1.38 s due to reordering. Decoding latency decreases with ratio in all three methods, and GPU memory follows the same trend. All methods converge toward total latencies of 4.58--5.54 s at their most aggressive settings, approaching the plain baseline. As established in Finding 4, however, generation performance has already converged to the no-context level at those ratios.

\begin{tcolorbox}[colback=gray!10, colframe=gray!50, boxrule=0.5pt, left=4pt, right=4pt, top=3pt, bottom=3pt]
\textbf{Finding 5.} All three paradigms reduce inference cost relative to full-context decoding across all evaluated ratios. T2I achieves the most pronounced improvement, with both compression and decoding costs decreasing monotonically to near-without-context levels at $128\times$. T2V compression overhead is small and ratio-independent; decoding latency decreases gradually, with all variants remaining well below full-context cost throughout. T2T compression overhead is method-dependent but ratio-stable; all methods approach without-context latency at aggressive compression settings.
\end{tcolorbox}

\section{Discussion}
\label{sec:discussion}

\subsection{Compression Fidelity}

\begin{figure*}[htbp]
    \centering
    \includegraphics[width=\textwidth]{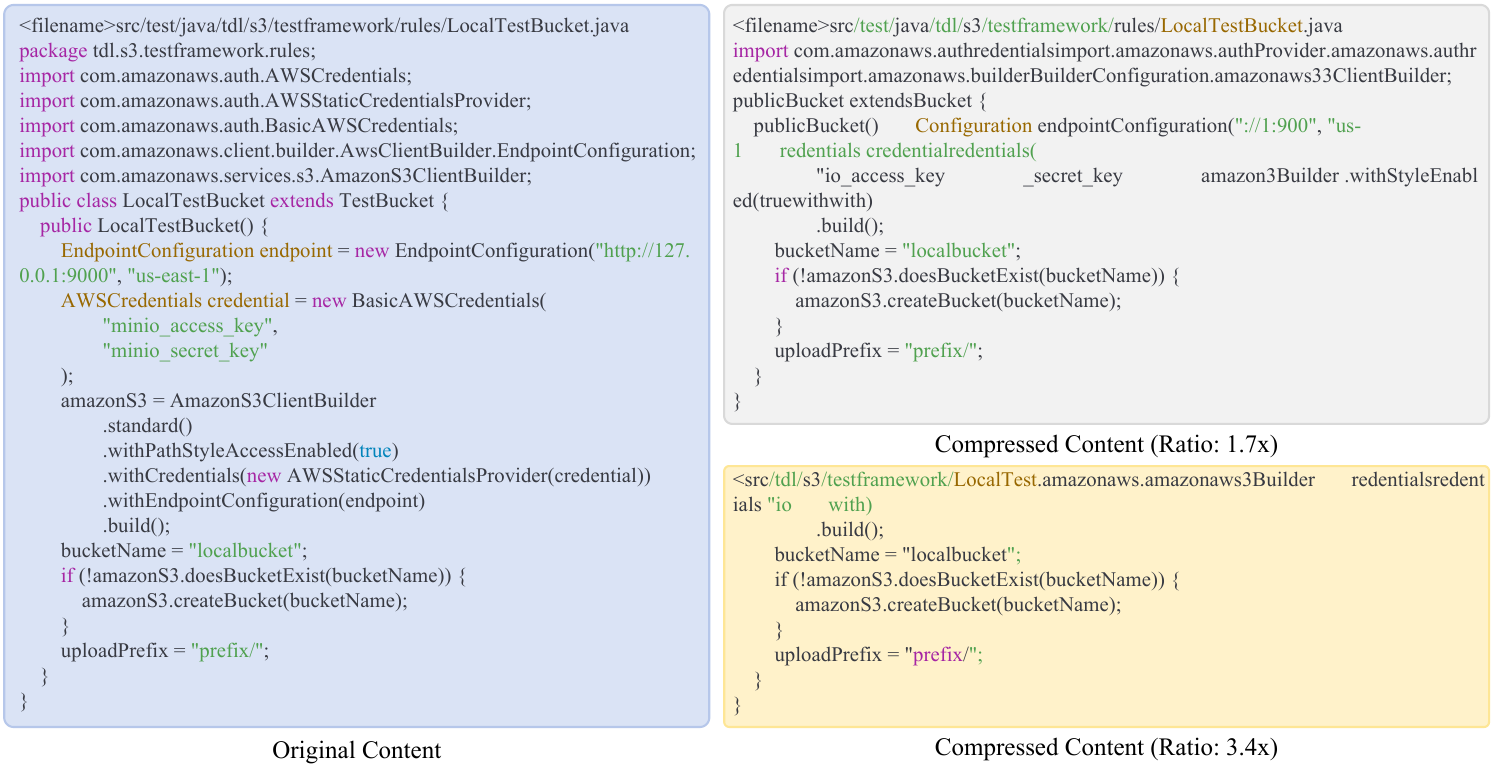}
    \caption{T2T (LLMLingua) compression applied to a representative Java source file. Left: original content. Middle and right: compressed outputs at 1.7$\times$ and 3.4$\times$ ratios, respectively.}
    \label{fig:case_study}
\end{figure*}

\begin{table}[htbp]
\centering
\caption{Text recovery results (BLEU and ES) for T2I and T2V across compression ratios, evaluated on 100 code files sampled from StarCoderData.}
\vspace{-6pt}
\renewcommand{\arraystretch}{0.8}
\label{tab:t2i_t2v_combined}
\begin{tabular}{lcccc}
\toprule
\multirow{2}{*}{\textbf{Ratio}} & \multicolumn{2}{c}{\textbf{Text-to-Image}} & \multicolumn{2}{c}{\textbf{Text-to-Vector}} \\
\cmidrule(r){2-3} \cmidrule(l){4-5}
& \textbf{BLEU} & \textbf{ES} & \textbf{BLEU} & \textbf{ES} \\
\midrule
2.0$\times$   & \textbf{63.53} & \textbf{82.50} & --    & --    \\
4.0$\times$   & \underline{22.80} & \underline{46.38} & \textbf{84.12} & \textbf{91.57} \\
8.0$\times$   & 4.59  & 24.85 & \underline{66.99} & \underline{80.49} \\
16.0$\times$  & 0.57  & 18.48 & 53.66 & 69.83 \\
32.0$\times$  & 0.35  & 15.49 & 35.55 & 56.87 \\
64.0$\times$  & 0.23  & 10.42 & 28.34 & 50.60 \\
128.0$\times$ & 0.18  & 7.45  & 16.51 & 41.09 \\
\bottomrule
\end{tabular}
\vspace{-6pt}
\renewcommand{\arraystretch}{1}
\end{table}
% \begin{table}[t]
% \centering
% \caption{Comparison of Te$\times$t2Image and Te$\times$t2Vector.}
% \label{tab:image_vector_comp}
% \small
% \begin{tabular}{lccc}
% \toprule
% \te$\times$tbf{Method} & \te$\times$tbf{Ratio} & \te$\times$tbf{BLEU} & \te$\times$tbf{ES} \\
% \midrule
% \multirow{7}{*}{Te$\times$t2Image} & 2.0$\times$ & 63.53\% & 82.50\% \\
%                             & 4.0$\times$ & 22.80\% & 46.38\% \\
%                             & 8.0$\times$ & 4.59\% & 24.85\% \\
%                             & 16.0$\times$ & 0.57\% & 18.48\% \\
%                             & 32.0$\times$ & 0.35\% & 15.49\% \\
%                             & 64.0$\times$ & 0.23\% & 10.42\% \\
%                             & 128.0$\times$ & 0.18\% & 7.45\% \\
% \midrule
% \multirow{6}{*}{Te$\times$t2Vector} & 4.0$\times$ & 84.12\% & 91.57\% \\
%                              & 8.0$\times$ & 66.99\% & 80.49\% \\
%                              & 16.0$\times$ & 53.66\% & 69.83\% \\
%                              & 32.0$\times$ & 35.55\% & 56.87\% \\
%                              & 64.0$\times$ & 28.34\% & 50.60\% \\
%                              & 128.0$\times$ & 16.51\% & 41.09\% \\
% \bottomrule
% \end{tabular}
% \end{table}

The three paradigms produce qualitatively different surrogate representations, which determines how their compression fidelity can be characterized.

\textbf{T2T} outputs human-readable compressed text, enabling direct qualitative inspection.
\autoref{fig:case_study} shows LLMLingua applied to a Java file at two ratios.
At 1.7$\times$, the output is nearly identical to the original---only minor whitespace and import consolidation occur.
At 3.4$\times$, several import statements and intermediate variable assignments are pruned while the core logic survives.
This reveals a structural weakness of perplexity-based scoring: imports and type annotations are linguistically predictable and thus receive low importance scores, yet they encode the cross-file dependencies that generation tasks rely on---consistent with the performance collapse observed at higher ratios in RQ2.

\textbf{T2I and T2V} produce non-human-readable representations (patch embeddings and latent memory vectors), making qualitative analysis infeasible.
We instead evaluate their information retention via a \textit{text recovery} task: reconstruct the original source code from the compressed representation and measure BLEU and ES against the ground truth.
We sampled 100 code files from StarCoderData~\cite{li2023starcodersourceyou}; results are shown in \autoref{tab:t2i_t2v_combined}.
T2V preserves substantially more information at every ratio---at 4$\times$, BLEU\,=\,84.12 / ES\,=\,91.57 versus T2I's 22.80 / 46.38; at 128$\times$, T2V still yields 16.51 / 41.09 while T2I collapses to 0.18 / 7.45.
This contrast reflects the core difference between the two paradigms: T2V's learned bottleneck actively distills task-relevant token structure, whereas T2I's rasterization distributes all content uniformly across pixels and irreversibly loses fine-grained lexical signals under aggressive downscaling---providing a direct explanation for the downstream performance gaps observed in RQ1 and RQ2.

% Since T2T outputs remain in the discrete token space, its fidelity loss is directly observable from the pruned sequences and has been fully characterized in RQ1 and RQ2; we therefore focus this analysis on T2I and T2V, whose patch embeddings and latent memory vectors are not human-readable. To quantify information retention, we evaluate a text recovery task on 100 code files sampled from StarCoderData~\cite{li2023starcodersourceyou}: the original source code is reconstructed from the compressed representation and scored against the ground truth using BLEU and ES (\autoref{tab:t2i_t2v_combined}). T2V retains substantially more information at every ratio: at 4$\times$, BLEU\,=\,84.12 and ES\,=\,91.57, compared to 22.80 and 46.38 for T2I; at 8$\times$, T2I drops to BLEU\,=\,4.59, while T2V remains at 66.99; at 128$\times$, T2V still recovers BLEU\,=\,16.51 / ES\,=\,41.09, whereas T2I collapses to 0.18 / 7.45. The sharp degradation of T2I beyond 4$\times$ is consistent with its uniform rendering: all tokens are allocated equal pixel area, so progressive downscaling erodes fine-grained lexical signals non-selectively, leaving the representation informationally empty at high ratios. T2V's learned bottleneck, by contrast, preserves structured content by explicitly optimising for reconstruction, which explains both its superior information retention and the downstream performance advantage observed in RQ1 and RQ2.

\subsection{Implications of Findings}

We summarize the implications of our study from both practitioners' and researchers' perspectives.

\textbf{For practitioners:}
Paradigm selection should be driven by task type, deployment constraints, and compression budget, as these factors interact in ways that no single method addresses optimally.
\begin{enumerate}
    \item \textit{When output performance is the primary requirement}, particularly for generation tasks, latent-vector compression (text-to-vector, T2V) is the recommended choice: it can surpass full-context performance and remains stable across a wide compression range ($4\times$ to $128\times$) with predictable, ratio-independent efficiency overhead. The key cost is that a task-specific compression module must be trained per base model, though the base LLM itself remains frozen throughout.

    \item \textit{When training-free deployment is required}, two options are available depending on model type. Visual compression (text-to-image, T2I) applies to vision-language models and is effective for completion tasks at moderate ratios; it should be avoided for generation tasks, as uniform rendering loses cross-file relational structure regardless of compression ratio. Token-filtering compression (text-to-text, T2T) is model-agnostic and requires no modification to the base LLM, making it the most portable option; however, it should be limited to mild ratios and larger models, as structurally critical tokens degrade rapidly under aggressive compression.

    \item \textit{When throughput or latency is the primary constraint}, T2I offers the most favorable progressive latency scaling and suits throughput-sensitive deployments; T2V provides stable, ratio-independent overhead well-suited to predictable serving environments.
\end{enumerate}

\textbf{For researchers:}
Our findings reveal several open challenges specific to programming language structure that do not arise in natural language settings, pointing to concrete directions for future work.
\begin{enumerate}
    \item T2V's supra-context performance indicates that repository context contains substantial redundancy that standard full-context decoding does not resolve. Researchers should investigate task-adaptive or query-conditioned memory allocation strategies to further improve the selectivity of the latent bottleneck.
    \item Perplexity-based scoring is poorly aligned with code structure: tokens most critical for cross-file reasoning are linguistically predictable and thus assigned low importance. Compression methods for code should incorporate program-specific signals such as AST node types, control-flow reachability, and cross-file dependency frequency.
    \item Visual compression currently allocates equal pixel area to all tokens, with no mechanism to prioritise structurally important regions. Layout-aware rendering strategies that assign higher resolution to function signatures and import blocks are a promising direction for improving fidelity under aggressive downscaling.
    \item The language-specific performance asymmetries observed across all three paradigms indicate that programming language syntax should be treated as a first-class signal in compression design rather than assumed to transfer uniformly across languages.
\end{enumerate}

\subsection{Threats to Validity}

\textbf{Hardware and Environment Variability.}
Efficiency metrics such as GPU memory and inference latency are sensitive to hardware configuration and software environment, and absolute values may not transfer directly to other setups. We mitigate this by running all methods and compression ratios on the same server under identical software environments, ensuring that the relative trends and cross-paradigm comparisons reported in RQ3 reflect differences in the compression operators rather than environmental noise.

\textbf{Potential Data Leakage.}
Since the pretraining corpora of the evaluated models are not publicly disclosed, we cannot fully rule out overlap with our evaluation instances. However, all methods are evaluated under identical prompts and experimental settings, so any leakage would affect all paradigms equally and would not distort the relative comparisons that are the focus of this study.

\section{Conclusion}
\label{sec:conclusion}

We present the first systematic empirical study of context compression for repository-level code intelligence, categorizing representative methods into three paradigms: Text-to-Vector (T2V), Text-to-Image (T2I), and Text-to-Text (T2T). We evaluate them on code completion and generation tasks across multiple model scales, jointly measuring task performance, ratio sensitivity, and deployment efficiency. Our study yields three principal findings. First, T2V surpasses full-context performance on both tasks, with the latent vectors acting as a selective filter on informationally sparse repository context rather than a mere compressor. Second, ratio sensitivity is paradigm-specific: T2V is stable across a wide compression range, T2I degrades with visual resolution due to non-selective rendering, and T2T's effective information loss outpaces its nominal token reduction. Third, T2I offers the most favorable latency scaling with compression ratio, while T2V incurs a fixed, predictable efficiency overhead; T2T provides latencies close to the context-free baseline only at ratios where performance has already collapsed. Together, these findings establish context compression as a viable and practically beneficial strategy for code LLMs, and surface open challenges specific to programming language structure that distinguish this setting from the natural language domain.

% \section*{Data Availability}
% Our source code and data are publicly available at~\cite{context_compress_2026}.

%%
%% The next two lines define the bibliography style to be used, and
%% the bibliography file.
\bibliographystyle{ACM-Reference-Format}
\bibliography{sample-base}

%%
%% If your work has an appendix, this is the place to put it.
\appendix

\end{document}